\documentclass[letter]{aa}  

\usepackage{graphicx}
\usepackage{natbib}
\usepackage[T1]{fontenc}
\bibpunct{(}{)}{;}{a}{}{,}

\newcommand{\micron}{$\mu$m}

\newcommand{\msun}{M$_{\odot}$}

\begin{document}

   \title{The Discovery of a Very Massive Star in W49\thanks{Based on observations made with ESO Telescopes at the La Silla Paranal Observatory under programme IDs 67.C-0514 and  073.D-0837) and on data acquired using the Large Binocular Telescope (LBT).}
%   \fnmsep\thanks{Based on data acquired using the Large Binocular Telescope (LBT). The LBT is an international collaboration among institutions in Germany, Italy, and the United States. LBT Corporation partners are LBT Beteiligungsgesellschaft, Germany, representing the Max Planck Society, the Astrophysical Institute Potsdam, and Heidelberg University; Istituto Nazionale di Astrofisica, Italy; The University of Arizona on behalf of the Arizona University system; The Ohio State University, and The Research Corporation, on behalf of The University of Notre Dame, University of Minnesota and University of Virginia}
 }

   \subtitle{}

   \author{S.-W. Wu\inst{1}\fnmsep\thanks{International Max Planck Research School for Astronomy and Cosmic Physics at the University of Heidelberg (IMPRS-HD)},
          A. Bik\inst{1}\fnmsep\inst{2},
          Th. Henning\inst{1}
          \and
          A. Pasquali\inst{3}
          \and 
          W. Brandner\inst{1}
          \and
          A. Stolte\inst{4}
                    }

   \institute{Max-Planck-Institut f\"{u}r Astronomie, K\"{o}nigstuhl 17, 69117 Heidelberg, Germany\\
	  \email{shiwei@mpia.de}
             \and
             Department of Astronomy, Stockholm University, AlbaNova University Centre, SE-106 91 Stockholm, Sweden
            \and
             Universit\"at Heidelberg, Zentrum f\"ur Astronomie, Astronomisches Recheninstitut, M\"onchhofstrasse 12-14, 69120 Heidelberg, Germany       
            \and
            Argelander Institut f\"ur Astronomie, Auf dem H\"ugel 71, 53121 Bonn, Germany
             }

\date{Received: 7 May 2014; accepted: 16 July 2014}

% \abstract{}{}{}{}{} 
% 5 {} token are mandatory
 
  \abstract
%  % context heading (optional)
%  % {} leave it empty if necessary  
   {Very massive stars (M > 100 \msun) are very rare objects, but have a strong influence on their environment. The formation of this kind of objects is of prime importance in star formation, but observationally still poorly constrained.}
%  % aims heading (mandatory)
   {We report on the identification of  a very massive star in the central cluster of the 
   star-forming region W49.}
%  % methods heading (mandatory)
   {We investigate near-infrared $K$-band spectroscopic observations of W49 from VLT/ISAAC together with  $JHK$ images obtained with NTT/SOFI and LBT/LUCI. We derive a spectral type of W49nr1, the brightest star in the dense core of the central cluster of W49.}
%  % results heading (mandatory)
   {On the basis of its $K$-band spectrum, W49nr1 is classified as an O2-3.5If* star with a $K$-band absolute magnitude of -6.27$\pm$0.10 mag. The effective temperature and bolometric correction are estimated from stars of similar spectral type. After comparison to the Geneva evolutionary models, we find an initial mass between 100 \msun\ and 180 \msun. Varying the extinction law results in a larger initial mass range of 90 - 250 \msun. %The present-day mass based on the luminosity is estimated to be......} %The age of W49nr1, however, is not well established.%  and an upper limit of 3 Myrs is derived. %The position of W49nr1 in the Hertzsprung Russel diagram suggests an upper age limit of  2 Myrs when compared with non-rotating isochrones, however, considering models with stellar rotation included, the age is estimated to be between 2 and 3 Myrs. 
   }
%  % conclusions heading (optional), leave it empty if necessary   With criterion on equivalent widths, it was classified as an O2-3If supergiant star.
   {}

   \maketitle
%
%________________________________________________________________

\section{Introduction}

Even though very massive stars (M > 100 \msun) are very rare, they have a strong influence on their environment via powerful winds and ionizing radiation, injecting large quantities of momentum and energy into the surrounding interstellar medium. Their fast evolution and the steep slope of the initial mass function (IMF) imply that one has to study the most massive star-forming regions to identify them. 

The formation mechanisms of very massive stars are by no means fully understood \citep{Krumholz14VMS}. % The two main scenarios explaining the formation of very massive stars involve accretion of stellar matter \citep[e.g.][]{Kuiper10, Kuiper:2012aa} and mergers of lower-mass stars in tight binaries \citep{Bonnell05}. 
For a long time, it was put in serious doubts whether these very massive stars could actually form at all. Observational evidence was presented suggesting an upper mass limit of 150 \msun~\citep{Figer05}. However,  recently \citet{Crowther2010aa} claimed the existence of   very massive stars up to 300 \msun, especially in and around young massive clusters, such as NGC 3603, the Arches cluster  and  R136  in the Large Magellanic Cloud. %{\bf The discovery of such objects would provide constraints on numerical simulation investigating the formation of massive stars (e.g. \citet{Kuiper:2010aa,Kuiper:2011aa}) } 

In this letter we present the discovery of a very massive star in one of the most luminous Galactic \ion{H}{II} regions: W49 (Fig \ref{W49image}). With dozens of OB-type stars in its core, W49 is one of the most important Galactic sites for studying the formation and evolution of  massive stars in the local universe \citep{Alves:2003aa,Homeier05}. Given its location in the plane of the Milky Way and a distance of 11.11 $^{+0.79}_{-0.69}$ kpc \citep{Zhang13}, W49 is optically obscured by intervening interstellar dust. This makes an optical identification and spectral classification of the stellar content close to impossible, leaving the near-infrared window (primarily $K$-band) for spectral classification of the highly obscured  stars.

Here, we report on the spectroscopic identification of a very massive star in W49, which we, hereafter, refer to as W49nr1. We first present our near-infrared observations (imaging and spectroscopy) of W49 (Sect.~2). The spectral features as well as the classification of W49nr1 are described in Sect.~3,  where we also derive its stellar parameters like effective temperature ($T_{\mbox{eff}}$), initial mass and age. Finally, we briefly discuss the implications of our results in Sect.~4 and end with  conclusions in Sect.~5. 

\section{Observation and data reduction}

A medium-resolution (R=10,000) $K$-band spectrum of W49nr1 was obtained with ISAAC mounted on Antu (UT1) of ESO's Very Large Telescope (VLT), Paranal, Chile.  $J$- and $H$-band  images were obtained with SOFI at the New Technology Telescope (NTT), La Silla, Chile and a $K$-band image was acquired with LUCI mounted on the Large Binocular Telescope (LBT), Mount Graham, Arizona.

\subsection{Observations}

SOFI $J$- and $H$-band imaging observations of W49 were performed on 2001, June 7 (PI: J. Alves) providing a $5' \times 5'$ field of view with $0.''288~pixel^{-1}$. The  data were taken with a DIT (detector integration time) of 6 s and NDIT (number of integrations) of 5 per saved frame. The number of exposures for $J$ and $H$-band are 20 and 15 respectively, which lead to a total exposure time of 600 s ($J$-band) and 450 s ($H$-band). The spatial resolution  is $ \sim 0.5-0.7''$.

The LUCI $K$-band data were taken on 2009, September 29 with the N3.75 camera, providing a $5' \times 5'$ field of view with $0.''12~pixel^{-1}$. The spatial resolution of the $K$ image is $ \sim 0.6-0.7''$.  The observations were taken with a DIT of 2 s and NDIT of 10. Forty-two frames were observed, resulting in a total exposure time of 840 s. Sky frames were taken at an offset positions centred at $\alpha (2000) = 19^h 08^m 35.8^s , \delta (2000) = +08^{\circ} 50' 52.7''$.

\begin{figure}
   \centering
   \includegraphics[width=\hsize]{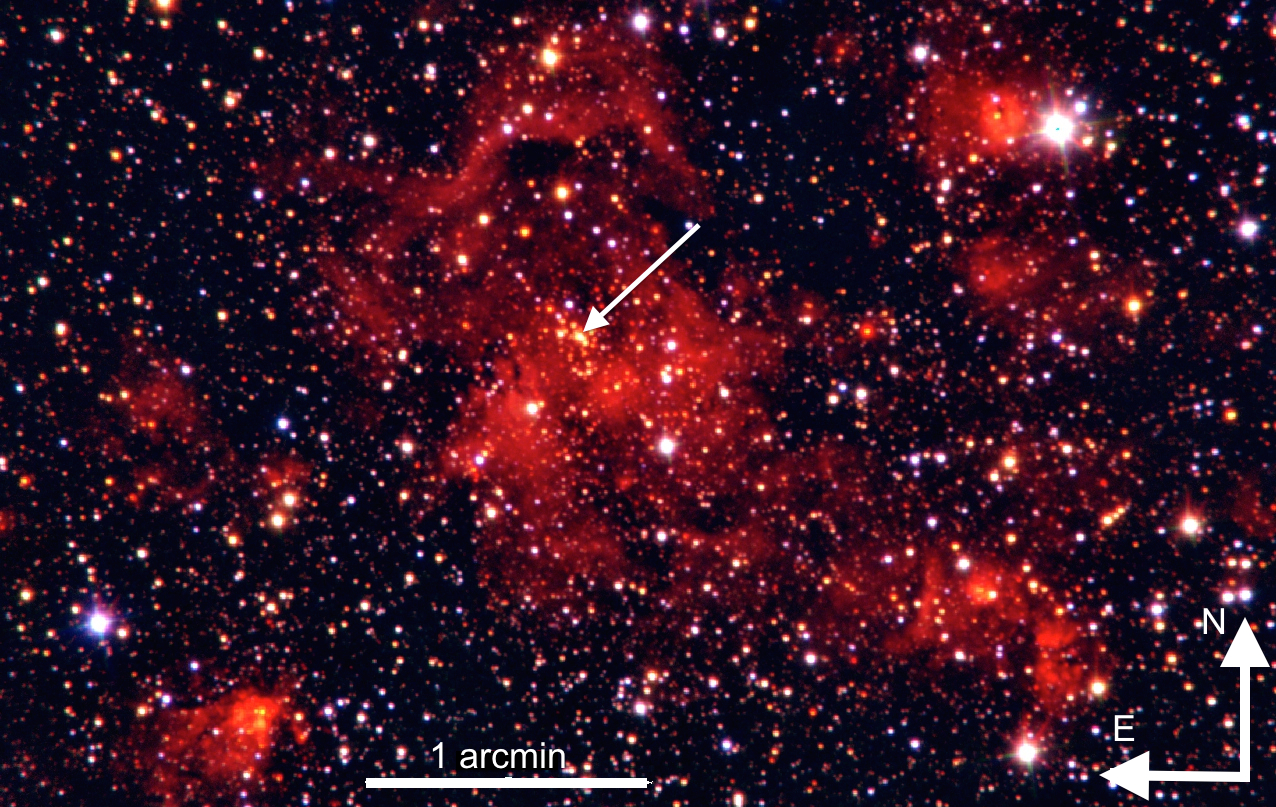}
      \caption{$JHK$ three color image of the central area of W49. The massive star W49nr1 is indicated with a white arrow.}
    \label{W49image}
\end{figure}

The most luminous star in the central cluster of W49, W49nr1 (Table \ref{tbl-1}), was observed with ISAAC in the $K$-band on 2004, August 6 (PI: J. Alves), with 3 exposures each with a DIT  of 300s. The  wavelength range covered by the spectrum is  2.08 $\mu$m to 2.20 $\mu$m. The sky frame and science frames were taken with an object-sky-object pattern, and the nodding offset between the two science frames was set to 20\arcsec. HR 6572, an A0V star, was used as the standard star to correct for the telluric features from the atmosphere. It was observed about one hour before the science frames, in the same wavelength range as the science observations  and with an integration time of 5 s. 

%($\alpha (2000) = 19^h 10^m 17.43^s, \delta (2000) = + 09^{\circ} 06' 20.93''$)

%Immediately after the science and standard star observations, arc lamp spectra and flat fields were taken to minimize the effects of flexure.% The resulting SNR (signal to noise ratio) of the spectrum is $\sim 90$.

\begin{figure}
   \centering
   \includegraphics[width=\hsize]{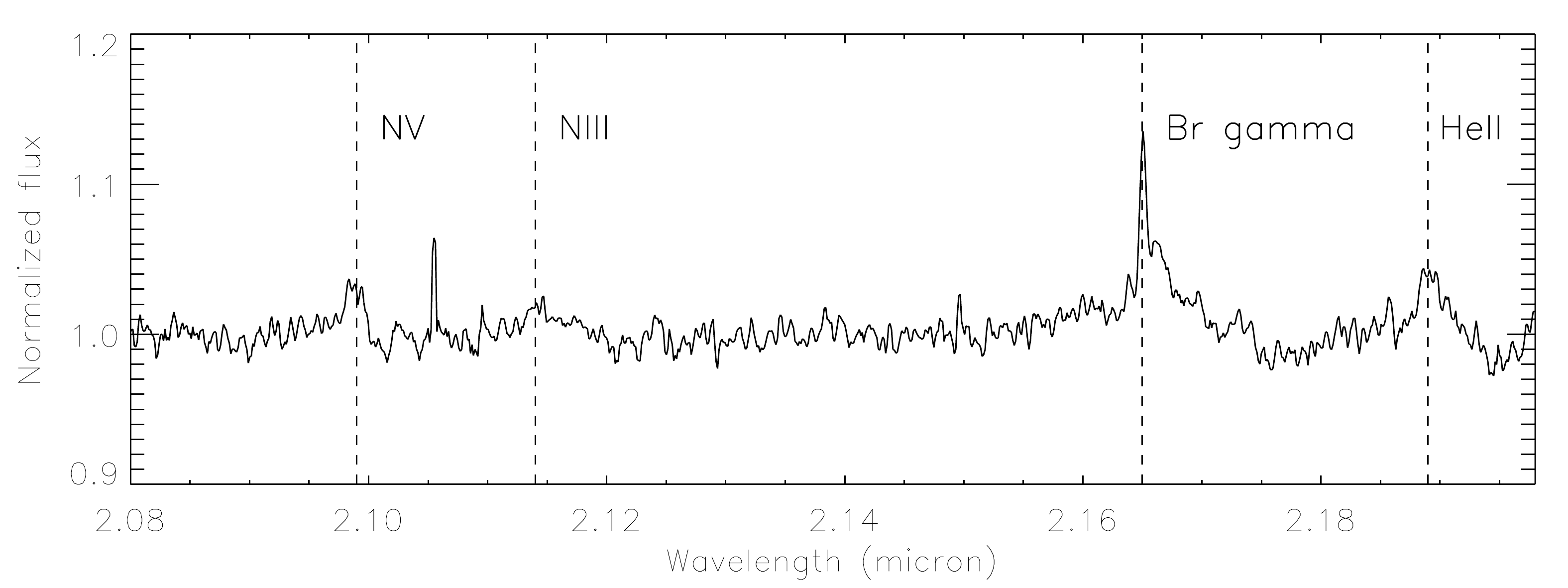}
      \caption{Normalised $K$-band spectrum of W49nr1 with the emission lines annotated.}
         \label{spectrum}
   \end{figure}

\subsection{Data reduction}
\subsubsection{Imaging}

The $J$- and $H$-band images were reduced using the ESO SOFI pipeline v1.5.2. and the $K$ data with standard IRAF routines \citep[see also][]{Pasquali:2011aa,Bik14}.  The  images were dark and flat field corrected. For the $JH$ data, a sky frame was created from the science frames by rejecting the brightest pixels while combining the frames in pixel coordinates. A sky frame for the $K$-band data was created by combining images taken at the offset position and rejecting the  3 lowest and  6 highest values. 

Photometry on the $JHK$ images was performed with  \texttt{DAOPHOT}  \citep{Stetson:1987aa} under IRAF. Stars are detected with \emph{daofind} with a threshold of 3$\sigma$ above the background. Aperture photometry was performed with  \emph{phot} in a radius of  ~(1-2) $\times$ the FWHM of the PSF. For each filter a reference PSF model was constructed by combining the PSF of at least 20 objects.  PSF-fitting photometry was performed with  \emph{allstar}, using the PSF model to fit all objects identified with a $3\sigma$ confidence level over the local background.

The $K$ image of W49 has severe nebular contamination strongly affecting the photometry of the point sources. 
To reduce the effect of the nebulosity in the $K$ image, we first removed the stars by means of PSF fitting. The residual frame, with all the stars subtracted, was then smoothed with a kernel of 12 pixels, resulting in a frame containing only the smooth nebular emission. This smoothed frame is subtracted from the original frame.  After that, we performed PSF  photometry on the nebular subtracted image, resulting in a more accurate photometry.
% with the nebular contribution removed.
W49nr1 is located in the center of a compact cluster and its photometry is affected by crowding from the neighbouring stars. 
To quantify the effect of the crowding, we performed aperture photometry at the position of  W49nr1 on the residuals in the psf-subtracted frame. This gives an error of 15.8, 6.4 and 8.6 \% for $J$, $H$ and $K$ respectively.

Finally, we cross-matched the obtained catalogs for each filter to identify the sources detected in more than one band.  We calibrated the SOFI and LUCI photometry with 2MASS \citep{Skrutskie:2006aa}. The final calibration resulted in errors in the zero points of 0.0063, 0.0071 and 0.0055 mag for $J$, $H$ and $K$-band respectively. We did not find a color dependence of the derived zero points.  The final errors of the $JHK$ photometry, as listed in Table~\ref{tbl-1}, are a combination of the photometry uncertainty, errors in the zero points and the errors due to crowding.

%After that we derived the magnitude difference between our photometry and the 2MASS magnitude then check whether this difference depend on the color of the sources. As a matter of fact, we found that the differences are almost constants and they do not change with magnitudes or colors. So we add the constants to our photometry and got the absolute magnitudes in the three band.

\begin{table}
\caption{Observed and derived properties of W49nr1}             % title of Table
\label{tbl-1}      % is used to refer this table in the text
\centering                          % used for centering table
\begin{tabular}{l r}        % centered columns (4 columns)
\hline\hline                 % inserts double horizontal lines
$\alpha$(J2000) (h m s) & 19:10:17.43 \\ 
$\delta$(J2000) ($^\circ$\ \arcmin\  \arcsec) & +9:06:20.93 \\
\hline
$J$ (mag) & 16.57$\pm0.18$ \\
$H$(mag) & 13.47$\pm0.12$ \\
$K$ (mag) &11.93$\pm0.10$ \\
\hline
EW(Br$\gamma$) (\AA) & 8.2 $\pm$ 1.7 \\
EW(\ion{He}{II}) (\AA) & 2.4 $\pm$ 0.7 \\
EW(\ion{N}{III}) (\AA) & 2.3 $\pm$ 1.0 \\
EW(\ion{N}{V}) (\AA) & 2.6 $\pm$ 0.9 \\
\hline
Spectral type & O2-3.5If* \\
$T_{\mbox{eff}}$ (K) & 40,000 -- 50,000 \\
BC (mag)	& -5.2 -- -4.55 \\
\hline
$A_{\mbox{K}}$ (mag) & 2.9\tablefootmark{a}/2.6 - 3.5\tablefootmark{b}\\
Initial mass (\msun) & 100 -- 180\tablefootmark{a}/90 -- 250\tablefootmark{b} \\
Luminosity (L/L$_{\sun}$) & 1.7 - 3.1$\times 10^6$\tablefootmark{a}/ 1.2 - 4.9$\times 10^6$\tablefootmark{b}\\
\hline
\end{tabular}
\tablefoot{
\tablefoottext{a} {With extinction law of \citet{Indebetouw05}.}
\tablefoottext{b} {Considering other extinction laws (see text).}
}
\end{table}

\subsubsection{Spectroscopy}
The ISAAC observations of W49nr1 were reduced using standard IRAF routines. The wavelength calibration was performed using the Xe and Ar arc frames. After the flat fielding and  wavelength calibration, the sky was removed by subtracting the frames taken at the A and B  nodding position. The spectra were extracted using \emph{doslit}  and the different exposures are combined to one final spectrum.  To remove the narrow $Br\gamma$ emission from the diffuse nebular emission surrounding the cluster, the background was estimated using a Legendre function, sampling a region close to the star, and subtracted from the spectrum.

The spectrum of the telluric standard star was reduced in the same way as the spectrum of W49nr1. Before correcting the spectrum of W49nr1 with the standard star, the Br$\gamma$ line of the standard star was removed by fitting a Lorentzian profile. The resulting atmospheric transmission spectrum was used to correct the science spectrum for telluric absorption using the  IRAF task \emph{telluric}. The signal-to-noise ratio (SNR) of the final spectrum is $\sim$90.

\section{Results}

\subsection{Spectral classification of W49nr1}

The final, normalized $K$-band spectrum of W49nr1 is shown in Fig.~\ref{spectrum}. The spectrum is dominated by broad emission lines of Br$\gamma$ (2.16 \micron), \ion{He}{II} (2.189 \micron), \ion{N}{III} (2.116 \micron) and \ion{N}{V} (2.10 \micron). The narrow emission component of $Br\gamma$ is a residual of the nebular subtraction. The \ion{He}{II} and \ion{N}{V} lines are indicative of an early spectral type \citep{Hanson05}. The broad emission profiles imply an origin in the stellar wind. These properties suggest similarities with the spectral classes  O2-3.5If*, O2-3.5If*/WN5-7 ("slash" stars) and WN5-7 stars  \citep{Crowther:2011aa}. The sum of the equivalent widths (EWs) of Br$\gamma$ and \ion{He}{II} can be used as a discriminator between these classes. For the WN5-7 stars, the summed EWs are expected to be above 70 \AA, while O2-3.5If* stars have a total EW between 2 and 20 \AA, with the "slash" stars lying in between. The total EW of both lines for W49nr1 (Table \ref{tbl-1}) is (10.6 $\pm$ 1.8 \AA), resulting in a classification of W49nr1 as O2-3.5If*.

%\citet{Crowther:2011aa} designed an optical and near-infrared classification system for these very massive stars.  They identify three different spectral classes using the H$\beta$\ line as a discriminating diagnostic. Because of the relatively small sample of stars with high-quality  $K$-band spectra, the divisions between Of, Of/WN and WN stars are less definitive than from the optical spectra.  

%\citet{Crowther:2011aa} designed a criterion based on Br$\gamma$ and \ion{He}{II} (2.189$\mu$m) in  the $K$-band to separate the 3 classes,  tracing the increase in the stellar wind density. 

 %An additional constraint on the spectral type is the \ion{N}{V} line at 2.10 \micron. This line is also detected by \citet{Hanson05} in an O31f* $K$-band spectrum, supporting our classification. 

\subsection{Hertzsprung-Russell diagram}

Based on the  classification of W49nr1 as an O2-3.5If* star,  we estimated $T_{\mbox{eff}}$ between 40,000 and 50,000 K and the bolometric correction (BC$_{K}$) between -5.2 and -4.55 mag adopting the derived values for an O2f*, O3I and an O4I star as representative for this class \citep{Crowther:2011aa}.   From our $HK$ photometry the absolute $K$-band  magnitude was derived to be -6.27 $\pm$  0.1 mag by assuming the distance of 11.11 kpc \citep{Zhang13}, applying the extinction law of \citet{Indebetouw05} and adopting the intrinsic color of $(H-K)=-0.1$ mag for O3I stars from \citet{Martins:2006aa}. After applying the BC$_{K}$, the bolometric magnitude of W49nr1 was derived to be between -11.47 and -10.82 mag, and  the corresponding bolometric luminosity between  $1.7\times 10^6$ and $3.1\times 10^6$ L$_{\sun}$. 

\begin{figure}
\begin{minipage}{0.242\textwidth}
\centering
   \includegraphics[width=\hsize]{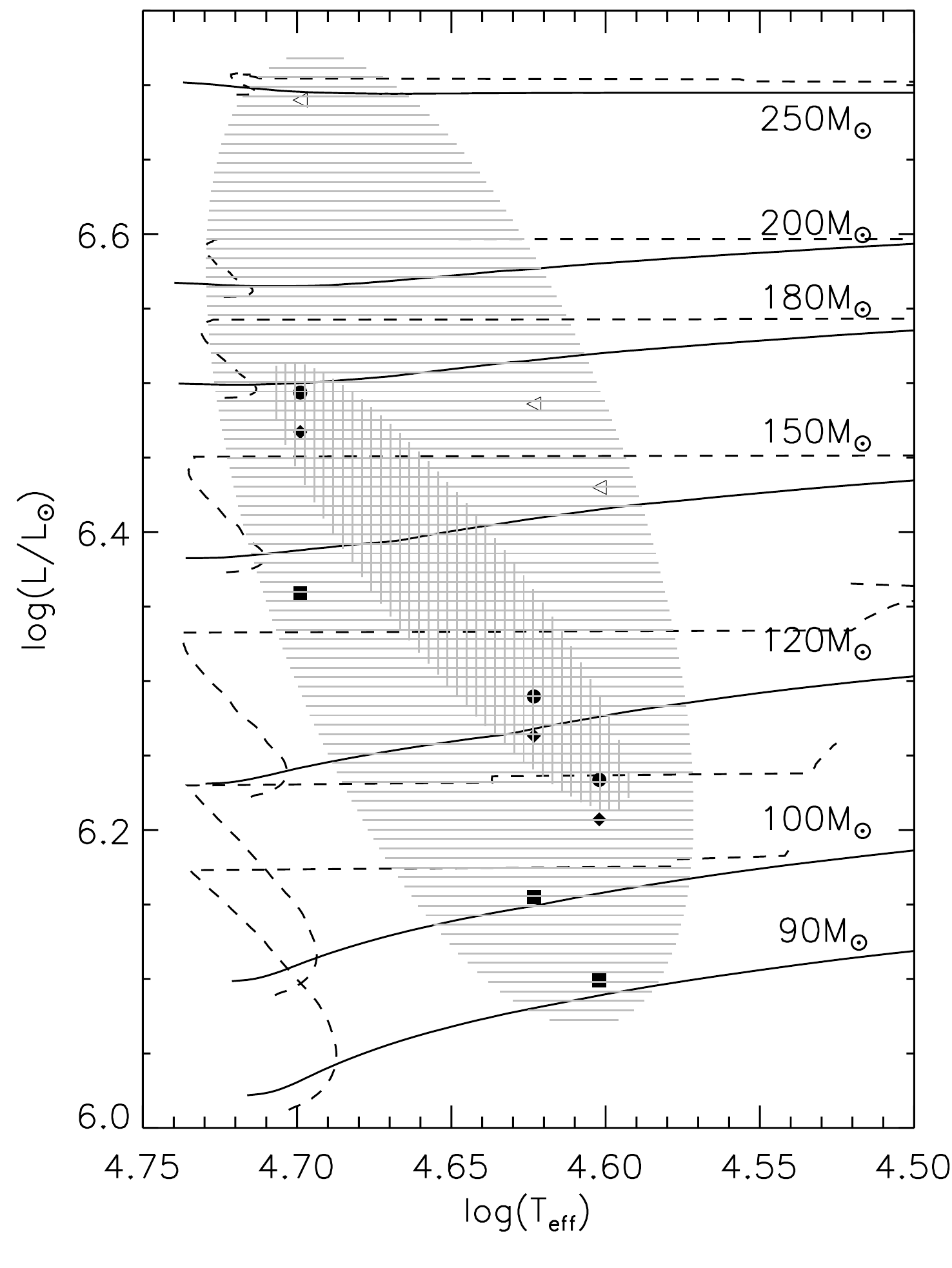}
\end{minipage}
\begin{minipage}{0.242\textwidth}
   \includegraphics[width=\hsize]{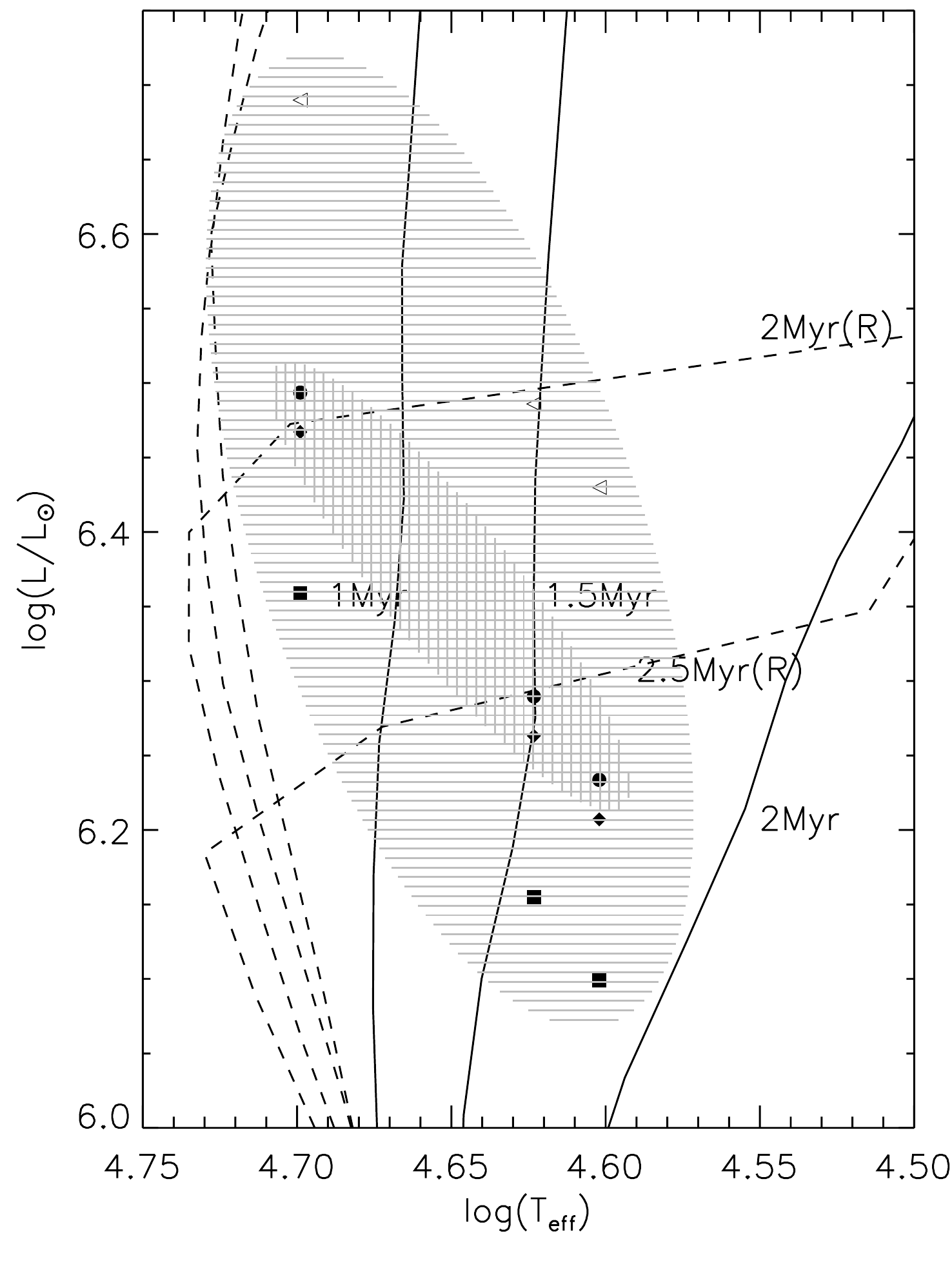}
\end{minipage}   
      \caption{HRD with the possible location of W49nr1 marked as a vertically hashed area and horizontally hashed, taking into account different extinction laws. The three black filled circles stand for an O2If*, an O3I and an O4I star respectively.  \emph{Left panel:} The Geneva evolution tracks \citep{Ekstrom:2012aa, Yusof:2013aa}  without stellar rotation (solid line) and with rotation (dashed line) for different masses are over plotted. \emph{Right panel:} The main sequence isochrones with different ages, again without stellar rotation (solid line) and with rotation (dashed line).}
    \label{HRD}
\end{figure}

We plotted the likely parameter space of W49nr1 in the Hertzsprung Russel diagram (HRD) as shown in Fig.~\ref{HRD}. As $T_{\mbox{eff}}$ and BC$_{K}$ are correlated, the likely location of W49nr1 is a diagonal ellipse. The possible locations of W49nr1 in the HRD was estimated by calculating the luminosity for the three spectral types in this class (O2If*, O3I and O4I), using their $T_{\mbox{eff}}$  and  corresponding BC$_{K}$.  To estimate the initial mass and age of W49nr1, the likely parameter space in the HRD  was compared with the Geneva theoretical stellar evolution models \citep{Ekstrom:2012aa,Yusof:2013aa}, using models with and without stellar rotation. 

From the evolutionary tracks, the initial mass of W49nr1 was estimated to be in the range between $\sim$110 and $\sim$180 \msun\ for models without rotation and between $\sim$100 and $\sim$170 \msun\ for models with rotation (Fig. \ref{HRD}, left panel, vertically hashed area).  While the initial mass estimate for W49nr1 is insensitive to rotation, the isochrones for the models with and without rotation for the same age are very different (Fig. \ref{HRD}, right panel). The position of W49nr1 suggests an upper age limit of $\sim$2 Myrs after comparison with the ``non-rotating'' isochrones, however, considering the models with rotation, an age between 2 and 3 Myrs is more likely.

As the extinction towards W49nr1 is high ($A_{\mbox{K}} = 2.9$ mag) the choice of the extinction law can have a large effect on the derived luminosity and therefore on its initial mass and age. To select the best fitting extinction laws, we applied a similar analysis to the color-color diagram of W49 as  \citet{Bik12} and found that the slopes of the  \citet{Cardelli89} and \citet{Roman07} laws were not consistent with the observations (Wu et al, in prep).

The extintion law of \citet{Indebetouw05} was the best fitting law, but also the slopes of \citet{Fitzpatrick99}, \citet{Nishiyama09} and \citet{Rieke85} are consistent with the observed colors. Taking into account all the 4 extinction laws, the estimated initial mass range widens to 90 - 250 \msun (see Fig. \ref{HRD}).

%If we assume the star formation processes in W49 are simultaneous, according to the four main sequence stars' positions in HRD (S.-W. Wu et al. 2014, in preparation), we could conclude the age of this cluster is younger than 2Myrs, thus rule out the rotation models and determine the upper limit on the mass of W49nr1 as 1.5 Myrs. If there exist an evolution sequence on the cluster...

\section{Discussion and future prospectives}

In this letter we report the discovery of a very massive star in the center of the main cluster in W49. In the following we discuss the uncertainties in the derivation of the stellar parameters and the implications for the properties of the central cluster in W49. We end with a suggestion for further characterization of this object.

\subsection{Stellar paramaters}

Our classification of W49nr1 depends on the empirical relation between the spectral type and the equivalent width of the emission lines as well as the calibration of $K$-band bolometric corrections for early-O stars based upon atmosphere models derived by \citet{Crowther:2011aa}. Due to the very few objects used in this study, it is hard to predict the uncertainty of this classification and a larger number of stars is needed to make this calibration more reliable.

The evolution of the very massive stars is mostly governed by their stellar wind and mass-loss rate. These input parameters for stellar evolution models add uncertainties to the estimated initial mass and age. As a comparison to the Geneva models we use the relation between the luminosity and the maximum stellar mass for homogeneous hydrogen burners \citep{Grafener11}, resulting in a present-day mass estimate of 110 - 175 \msun\ (and  95 - 250 \msun\ for taking into account all 4 extinction laws as discussed in Sect. 3.2).

The stellar rotation only plays an important part in estimating the age of W49nr1 from the HRD, as the rotational models predict longer time scales for the different evolutionary phases of the massive stars. The $K$-band spectrum is fully dominated by emission lines originating in the stellar wind, hence no estimate of the rotation can be made. High resolution spectroscopy of possible absorption lines to derive its rotation is key to understand the evolutionary status of this extreme star as well as the cluster. By monitoring the radial velocity of the emission lines, multi-epoch spectroscopy could probe for a possible binary nature.

\subsection{Cluster properties}
W49nr1 is located in the center of the compact central cluster in W49 (Fig. \ref{W49image}), and thus supports the theoretical expectation of rapid dynamical mass segregation (e.g. \citet{Allison:2009aa}). \citet{Homeier05} estimate the mass of this cluster as $10^4$ \msun. This suggests that W49nr1 is located in an environment quite similar to other very massive stars located in \citep{Crowther2010aa}. It adds to the growing number of stars with initial masses at or above 150 \msun, suggesting the absence of a strict upper mass limit for massive stars as also suggested by numerical simulations \citep{Kuiper:2010aa,Kuiper:2011aa}.

Comparing the cluster mass and the derived stellar mass for W49nr1 to theoretical relations between cluster mass and mass of the most massive star \citep{Weidner10} shows that the presence of such a massive star is consistent with a normal initial mass function. A full study of the upper end of the IMF requires a spectroscopy classification of many more massive stars (Wu et al, in prep).

%X-Ray source associated with the WRstar or not? Compare the X-ray luminosity to the bolometric luminosity ratio with the canonical value excepted for single star $10^{-7}$. 
%
%Or high quality image of that region? If it is a close system, or the brightness difference between two stars is too large?(Then the secondary is not important)
%
%HST/NICMOS, or NACO?
%
%for the binary case......assume the magnitude difference, to get the masses of each star. See the samples with similar spectral type: WR20a \citep{Bonanos:2004aa, Rauw:2004aa} and WR21a \citep{Niemela:2008aa}.

\section{Conclusions}

In this letter we present  $JHK_s$ imaging and $K$-band spectroscopy observations of W49nr1, the brightest star in the central cluster of W49. According to classification criteria based on the equivalent widths of Br$\gamma$ and \ion{He}{II} given by \citet{Crowther:2011aa}, W49nr1 is classified as an O2-3.5If* star. We estimate the effective temperature to be between 40,000 and 50,000 K and the bolometric luminosity between $1.7\times 10^6$ and $3.1\times 10^6$ L$_{\sun}$. Comparison with the Geneva stellar evolutionary tracks suggests an initial mass range of 100 - 180 \msun\ in the case of a single star, relatively independent of rotational velocity. We study the effect of variations in the extinction law on the stellar parameters, resulting in a large initial mass range of 90 - 250 \msun. Estimates of the present day mass delivers similar values. The age depends severely on rotational velocity and can only be constrained to less than 3 Myrs. The next step will be a full spectroscopic modelling of the near-infrared spectrum of W49nr1 resulting in stricter constraints on the effective temperature and luminosity. Spectral modelling will allow us to identify possible absorption lines at other wavelengths, suitable for measuring the rotational velocity.

\begin{acknowledgements}
We thank the anonymous referee for helpful suggestions which have improved the paper significantly. We acknowledge Fabrice Martins and Adrianne Liermann for extensive discussions on the interpretation of the K-band spectrum.   A.B. acknowledges the hospitality of the Aspen Center for Physics, which is supported by the National Science Foundation Grant No. PHY-1066293.  
The LBT is an international collaboration among institutions in Germany, Italy, and the United States. LBT Corporation partners are LBT Beteiligungsgesellschaft, Germany, representing the Max Planck Society, the Astrophysical Institute Potsdam, and Heidelberg University; Istituto Nazionale di Astrofisica, Italy; The University of Arizona on behalf of the Arizona University system; The Ohio State University, and The Research Corporation, on behalf of The University of Notre Dame, University of Minnesota and University of Virginia
\end{acknowledgements}

%-------------------------------------------------------------------

\bibliographystyle{aa}
\bibliography{my_bib,arjanW49_bib}

\begin{thebibliography}{27}
\expandafter\ifx\csname natexlab\endcsname\relax\def\natexlab#1{#1}\fi

\bibitem[{{Allison} {et~al.}(2009){Allison}, {Goodwin}, {Parker}, {de Grijs},
  {Portegies Zwart}, \& {Kouwenhoven}}]{Allison:2009aa}
{Allison}, R.~J., {Goodwin}, S.~P., {Parker}, R.~J., {et~al.} 2009, \apjl, 700,
  L99

\bibitem[{{Alves} \& {Homeier}(2003)}]{Alves:2003aa}
{Alves}, J. \& {Homeier}, N. 2003, \apjl, 589, L45

\bibitem[{Bik {et~al.}(2012)Bik, Henning, Stolte, Brandner, Gouliermis,
  Gennaro, Pasquali, Rochau, Beuther, Ageorges, Seifert, Wang, \&
  Kudryavtseva}]{Bik12}
Bik, A., Henning, T., Stolte, A., {et~al.} 2012, \apj, 744, 87

\bibitem[{Bik {et~al.}(2014)Bik, Stolte, Gennaro, Brandner, Gouliermis,
  Hussmann, Tognelli, Rochau, Henning, Adamo, Beuther, Pasquali, \&
  Wang}]{Bik14}
Bik, A., Stolte, A., Gennaro, M., {et~al.} 2014, \aap, 561, 12

\bibitem[{Cardelli {et~al.}(1989)Cardelli, Clayton, \& Mathis}]{Cardelli89}
Cardelli, J.~A., Clayton, G.~C., \& Mathis, J.~S. 1989, \apj, 345, 245

\bibitem[{{Crowther} {et~al.}(2010){Crowther}, {Schnurr}, {Hirschi}, {Yusof},
  {Parker}, {Goodwin}, \& {Kassim}}]{Crowther2010aa}
{Crowther}, P.~A., {Schnurr}, O., {Hirschi}, R., {et~al.} 2010, \mnras, 408,
  731

\bibitem[{{Crowther} \& {Walborn}(2011)}]{Crowther:2011aa}
{Crowther}, P.~A. \& {Walborn}, N.~R. 2011, \mnras, 416, 1311

\bibitem[{{Ekstr{\"o}m} {et~al.}(2012){Ekstr{\"o}m}, {Georgy}, {Eggenberger},
  {Meynet}, {Mowlavi}, {Wyttenbach}, {Granada}, {Decressin}, {Hirschi},
  {Frischknecht}, {Charbonnel}, \& {Maeder}}]{Ekstrom:2012aa}
{Ekstr{\"o}m}, S., {Georgy}, C., {Eggenberger}, P., {et~al.} 2012, \aap, 537,
  A146

\bibitem[{Figer(2005)}]{Figer05}
Figer, D.~F. 2005, Nature, 434, 192

\bibitem[{Fitzpatrick(1999)}]{Fitzpatrick99}
Fitzpatrick, E.~L. 1999, \pasp, 111, 63

\bibitem[{Gr{\"a}fener {et~al.}(2011)Gr{\"a}fener, Vink, de~Koter, \&
  Langer}]{Grafener11}
Gr{\"a}fener, G., Vink, J.~S., de~Koter, A., \& Langer, N. 2011, \aap, 535, 56

\bibitem[{Hanson {et~al.}(2005)Hanson, Kudritzki, Kenworthy, Puls, \&
  Tokunaga}]{Hanson05}
Hanson, M.~M., Kudritzki, R.-P., Kenworthy, M.~A., Puls, J., \& Tokunaga, A.~T.
  2005, \apjs, 161, 154

\bibitem[{Homeier \& Alves(2005)}]{Homeier05}
Homeier, N.~L. \& Alves, J. 2005, \aap, 430, 481

\bibitem[{Indebetouw {et~al.}(2005)Indebetouw, Mathis, Babler, Meade, Watson,
  Whitney, Wolff, Wolfire, Cohen, Bania, Benjamin, Clemens, Dickey, Jackson,
  Kobulnicky, Marston, Mercer, Stauffer, Stolovy, \& Churchwell}]{Indebetouw05}
Indebetouw, R., Mathis, J.~S., Babler, B.~L., {et~al.} 2005, \apj, 619, 931

\bibitem[{Krumholz(2014)}]{Krumholz14VMS}
Krumholz, M.~R. 2014, in Very Massive Stars in the Local Universe, 3417

\bibitem[{{Kuiper} {et~al.}(2010){Kuiper}, {Klahr}, {Beuther}, \&
  {Henning}}]{Kuiper:2010aa}
{Kuiper}, R., {Klahr}, H., {Beuther}, H., \& {Henning}, T. 2010, \apj, 722,
  1556

\bibitem[{{Kuiper} {et~al.}(2011){Kuiper}, {Klahr}, {Beuther}, \&
  {Henning}}]{Kuiper:2011aa}
{Kuiper}, R., {Klahr}, H., {Beuther}, H., \& {Henning}, T. 2011, \apj, 732, 20

\bibitem[{{Martins} \& {Plez}(2006)}]{Martins:2006aa}
{Martins}, F. \& {Plez}, B. 2006, \aap, 457, 637

\bibitem[{Nishiyama {et~al.}(2009)Nishiyama, Tamura, Hatano, Kato, Tanab{\'e},
  Sugitani, \& Nagata}]{Nishiyama09}
Nishiyama, S., Tamura, M., Hatano, H., {et~al.} 2009, \apj, 696, 1407

\bibitem[{{Pasquali} {et~al.}(2011){Pasquali}, {Bik}, {Zibetti}, {Ageorges},
  {Seifert}, {Brandner}, {Rix}, {J{\"u}tte}, {Knierim}, {Buschkamp}, {Feiz},
  {Gemperlein}, {Germeroth}, {Hofmann}, {Laun}, {Lederer}, {Lehmitz}, {Lenzen},
  {Mall}, {Mandel}, {M{\"u}ller}, {Naranjo}, {Polsterer}, {Quirrenbach},
  {Sch{\"a}ffner}, {Storz}, \& {Weiser}}]{Pasquali:2011aa}
{Pasquali}, A., {Bik}, A., {Zibetti}, S., {et~al.} 2011, \aj, 141, 132

\bibitem[{Rieke \& Lebofsky(1985)}]{Rieke85}
Rieke, G.~H. \& Lebofsky, M.~J. 1985, \apj, 288, 618

\bibitem[{Rom{\'a}n-Z{\'u}{\~n}iga {et~al.}(2007)Rom{\'a}n-Z{\'u}{\~n}iga,
  Lada, Muench, \& Alves}]{Roman07}
Rom{\'a}n-Z{\'u}{\~n}iga, C.~G., Lada, C.~J., Muench, A., \& Alves, J.~F. 2007,
  \apj, 664, 357

\bibitem[{{Skrutskie} {et~al.}(2006){Skrutskie}, {Cutri}, {Stiening},
  {Weinberg}, {Schneider}, {Carpenter}, {Beichman}, {Capps}, {Chester},
  {Elias}, {Huchra}, {Liebert}, {Lonsdale}, {Monet}, {Price}, {Seitzer},
  {Jarrett}, {Kirkpatrick}, {Gizis}, {Howard}, {Evans}, {Fowler}, {Fullmer},
  {Hurt}, {Light}, {Kopan}, {Marsh}, {McCallon}, {Tam}, {Van Dyk}, \&
  {Wheelock}}]{Skrutskie:2006aa}
{Skrutskie}, M.~F., {Cutri}, R.~M., {Stiening}, R., {et~al.} 2006, \aj, 131,
  1163

\bibitem[{{Stetson}(1987)}]{Stetson:1987aa}
{Stetson}, P.~B. 1987, \pasp, 99, 191

\bibitem[{Weidner {et~al.}(2010)Weidner, Kroupa, \& Bonnell}]{Weidner10}
Weidner, C., Kroupa, P., \& Bonnell, I. A.~D. 2010, \mnras, 401, 275

\bibitem[{{Yusof} {et~al.}(2013){Yusof}, {Hirschi}, {Meynet}, {Crowther},
  {Ekstr{\"o}m}, {Frischknecht}, {Georgy}, {Abu Kassim}, \&
  {Schnurr}}]{Yusof:2013aa}
{Yusof}, N., {Hirschi}, R., {Meynet}, G., {et~al.} 2013, \mnras, 433, 1114

\bibitem[{Zhang {et~al.}(2013)Zhang, Reid, Menten, Zheng, Brunthaler, Dame, \&
  Xu}]{Zhang13}
Zhang, B., Reid, M.~J., Menten, K.~M., {et~al.} 2013, \apj, 775, 79

\end{thebibliography}

\end{document}